\documentclass[prb,twocolumn,reprint,superscriptaddress]{revtex4-2}
\usepackage[utf8]{inputenc}
\usepackage{amsmath}
\usepackage{amssymb}
\usepackage{bm}
\usepackage{graphicx}
\usepackage{graphics}
\usepackage{color}
\usepackage{textcomp}
\usepackage[bookmarks=false]{hyperref}
\usepackage[usenames,dvipsnames]{xcolor}
\usepackage{subfigure}
\usepackage[super]{nth}
\def\be{\begin{equation}}
\def\ee{\end{equation}}
\def\ba{\begin{eqnarray}}
\def\ea{\end{eqnarray}}

\def\pet3{Pd$_x$ErTe$_3$}

\begin{document}


\title{Anomalous thermal transport and violation of  Wiedemann-Franz law in the critical regime of a charge density wave transition }


\author{Erik Kountz}
\affiliation{Stanford Institute for Materials and Energy Sciences,\\
SLAC National Accelerator Laboratory, 2575 Sand Hill Road, Menlo Park, CA 94025}
\affiliation{Geballe Laboratory for Advanced Materials, Stanford University, Stanford, CA 94305}
\affiliation{Department of Applied Physics, Stanford University, Stanford, CA 94305}

\author{Jiecheng Zhang}
\affiliation{Stanford Institute for Materials and Energy Sciences,\\
SLAC National Accelerator Laboratory, 2575 Sand Hill Road, Menlo Park, CA 94025}
\affiliation{Geballe Laboratory for Advanced Materials, Stanford University, Stanford, CA 94305}
\affiliation{Department of Applied Physics, Stanford University, Stanford, CA 94305}

\author{Joshua A. W. Straquadine}
\affiliation{Stanford Institute for Materials and Energy Sciences,\\
SLAC National Accelerator Laboratory, 2575 Sand Hill Road, Menlo Park, CA 94025}
\affiliation{Geballe Laboratory for Advanced Materials, Stanford University, Stanford, CA 94305}
\affiliation{Department of Applied Physics, Stanford University, Stanford, CA 94305}

\author{Anisha G. Singh}
\affiliation{Stanford Institute for Materials and Energy Sciences,\\
SLAC National Accelerator Laboratory, 2575 Sand Hill Road, Menlo Park, CA 94025}
\affiliation{Geballe Laboratory for Advanced Materials, Stanford University, Stanford, CA 94305}
\affiliation{Department of Applied Physics, Stanford University, Stanford, CA 94305}

\author{Maja D. Bachmann}
\affiliation{Stanford Institute for Materials and Energy Sciences,\\
SLAC National Accelerator Laboratory, 2575 Sand Hill Road, Menlo Park, CA 94025}
\affiliation{Geballe Laboratory for Advanced Materials, Stanford University, Stanford, CA 94305}
\affiliation{Department of Applied Physics, Stanford University, Stanford, CA 94305}

\author{Ian R. Fisher}
\affiliation{Stanford Institute for Materials and Energy Sciences,\\
SLAC National Accelerator Laboratory, 2575 Sand Hill Road, Menlo Park, CA 94025}
\affiliation{Geballe Laboratory for Advanced Materials, Stanford University, Stanford, CA 94305}
\affiliation{Department of Applied Physics, Stanford University, Stanford, CA 94305}

\author{Steven A. Kivelson}
\affiliation{Stanford Institute for Materials and Energy Sciences,\\
SLAC National Accelerator Laboratory, 2575 Sand Hill Road, Menlo Park, CA 94025}
\affiliation{Geballe Laboratory for Advanced Materials, Stanford University, Stanford, CA 94305}
\affiliation{Department of Physics, Stanford University, Stanford, CA 94305}

\author{Aharon Kapitulnik}
\affiliation{Stanford Institute for Materials and Energy Sciences,\\
SLAC National Accelerator Laboratory, 2575 Sand Hill Road, Menlo Park, CA 94025}
\affiliation{Geballe Laboratory for Advanced Materials, Stanford University, Stanford, CA 94305}
\affiliation{Department of Applied Physics, Stanford University, Stanford, CA 94305}
\affiliation{Department of Physics, Stanford University, Stanford, CA 94305}


\date{\today}

\begin{abstract}
ErTe$_3$ is studied as a model system to explore thermal 
transport in a layered charge density wave (CDW) material. We present data from thermal diffusivity, resistivity, and specific heat measurements: 
There is a sharp decrease in thermal conductivity both parallel and perpendicular to the primary CDW at the CDW transition temperature. At the same time, the resistivity changes more gradually.  Correspondingly,  while well above and below $T_c$, a consistent description of  the thermal transport applies with essentially independent electron and phonon contributions (estimated using the Wiedemann Franz law), 
in the critical regime  
no such description is possible; the observed behavior corresponds to a strongly coupled electron-phonon critical `soup.'
\end{abstract}


\maketitle


\section{Introduction}

Contrary to the standard paradigm  \cite{Peierls1955,Frohlich1954}, 
 in more than one spatial dimension, due to the  lack of perfect nesting,  charge density wave (CDW) order (unlike superconducting order) only emerges for interactions greater than a critical strength.  Generically, both strong electron-electron and electron-phonon interactions are involved in the ``mechanism.''  Among other things, the strong-coupling aspect is reflected in large values of the induced gap to $T_c$ ratios - in the focus material for the present study, 
 ErTe$_3$,  the gap associated with the primary CDW transition at $T_{CDW1} \approx 265~$K is $\Delta_1\approx 175~$meV, so that $2\Delta_1/k_BT_{CDW1}\approx 15$ \cite{Pfuner2010,Moore2010}. However,  especially since conventional notions of weakly interacting quasi-particles and well defined phonons give a good account of the normal state at $T > T_{CDW1}$ and of the low temperature physics deep in the ordered phase, it is conventional (as is indeed common more generally in considering classical critical phenomena in metals) to adopt a phenomenological approach in which the low energy quasi-particles and the bulk of the phonons are weakly coupled to the ``critical modes'' associated directly with the CDW transition.  This perspective is challenged by our present results.

Electrical and thermal transport measurements provide important information about the electronic structure and scattering processes in complex quantum materials. 
To the extent that the transport is dominated by weakly interacting (emergent) elementary excitations, the thermal conductivity can be expressed as the sum of an electronic and a phonon piece, $\kappa\approx \kappa_{el} + \kappa_{ph}$. Moreover, to the extent that the scattering processes are quasi-elastic, $\kappa_{el}$ is related to the electrical conductivity by the so-called Wiedemann-Franz (WF) law, i.e. the ratio between them, $\kappa_{el}/\sigma=L_0 T$ is determined by the universal constant,  $L_0=8k_B^2/\pi e^2\approx 2.44\times 10^{-8}$W$\Omega$K$^{-2}$. Observing this ratio gives information about how ``standard''  the  transport  is in a given electronic system,  with significant violations of the WF law plausibly indicating the breakdown of the quasiparticle description.

In this paper we examine electrical and thermal transport in the layered material ErTe$_3$, which has two successive CDW transitions at $T_{CDW1} \approx 265~$K and $T_{CDW2} \approx 160~$K.  Here $T_{CDW1}$ marks the onset of the ``primary'' CDW order with ordering vector $q_{CDW1}$ along the $c$-direction, while below $T_{CDW2}$ a  ``secondary'' orthogonal CDW component with $q_{CDW2}$ along the $a$-direction appears (where $a$ and $c$ are in-plane lattice parameters in the standard space group setting for this structure type).  Thus, despite the nearly tetragonal symmetry of the crystal ($a \sim c$), the phase at $T_{CDW1} > T > T_{CDW2}$ has unidirectional CDW order, while the low temperature
CDW is bidirectional, but with  inequivalent amplitudes in the two directions.  ErTe$_3$ is an ideal ``model system'' in that it is  stoichiometric and  hence can be synthesized with a high degree of crystalline perfection and little disorder (in the sense both of very low residual resistivity - $\rho(T) < 1~\mu\Omega$-cm at low $T$  
- and resolution limited Bragg peaks associated with the CDW order). It  boasts  broad metallic bands with a plasma frequency which, depending on the global fit, is estimated in the range of $2.5~$eV \cite{Pfuner2010} to $5.8~$eV \cite{Hu2011} in the CDW state.   Moreover, the effect of disorder can be explored systematically by Pd intercalation, as discussed in previous publications.\cite{Straquadine2019,Fang2019,Fang2020}

\begin{figure}[ht]
\centering
\subfigure{\label{cp}\includegraphics[width=0.95\columnwidth]{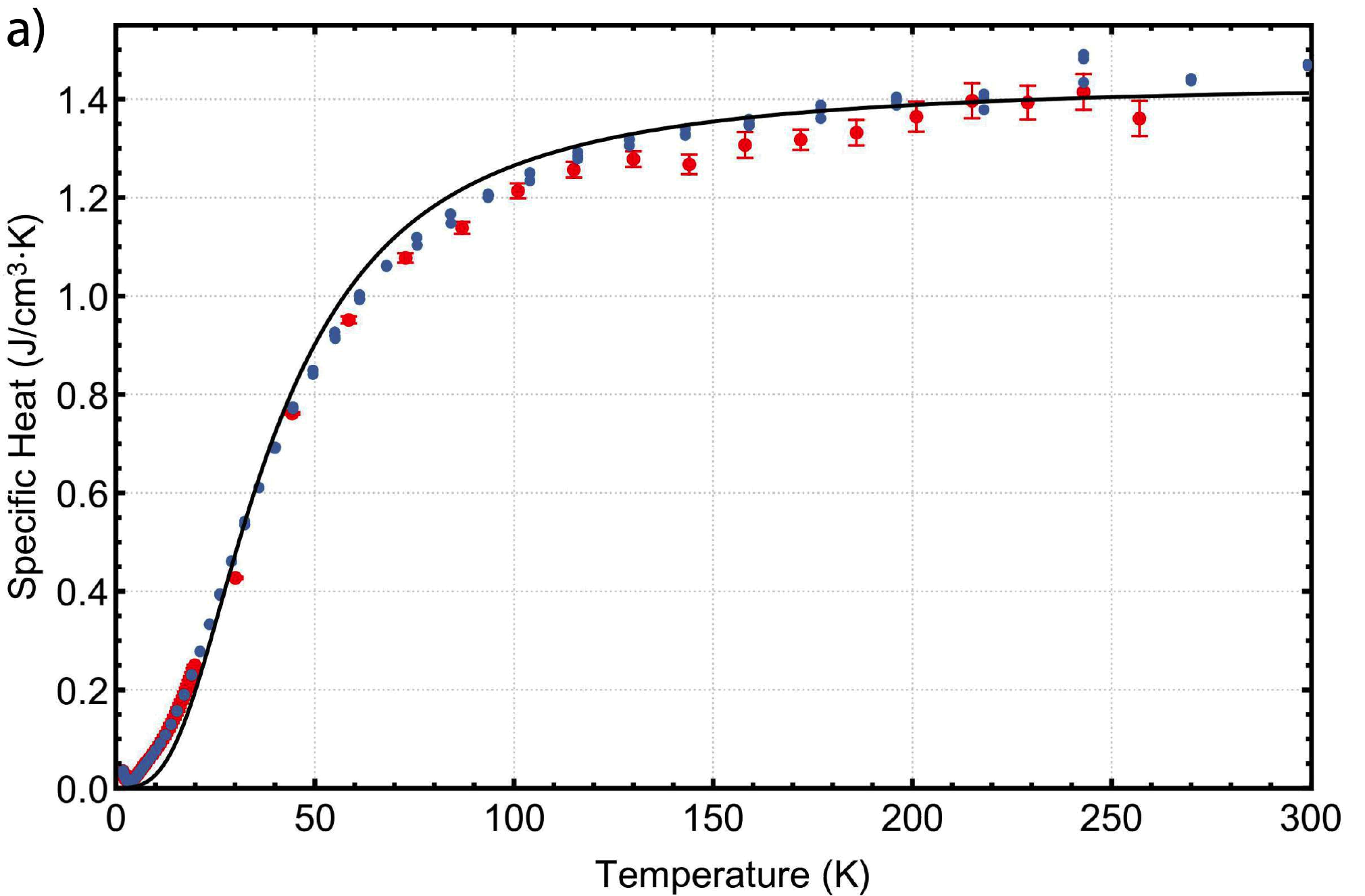} }
\subfigure{\label{res}\includegraphics[width=0.95\columnwidth]{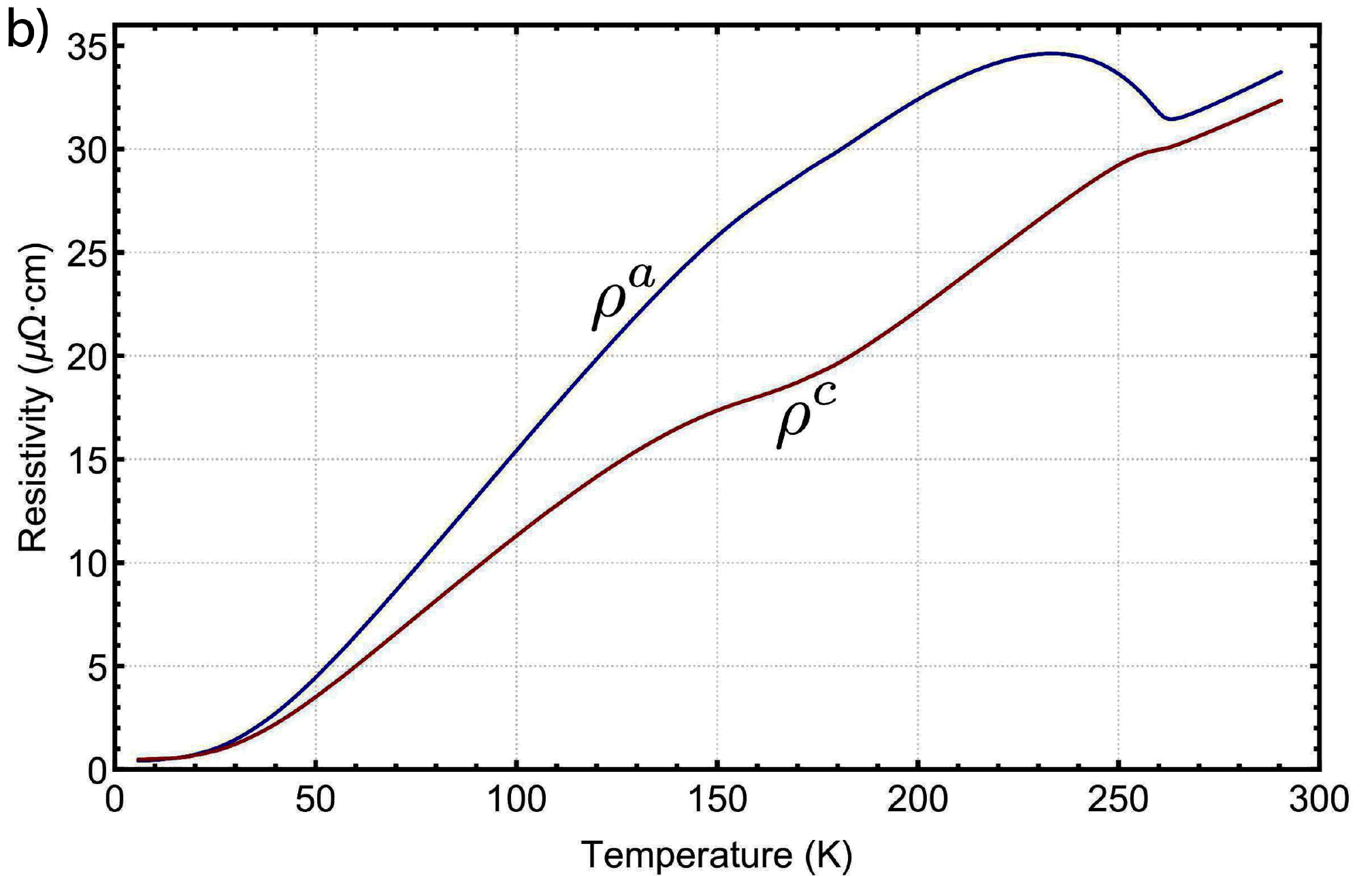} }
\caption{ (a) Specific heat of two different crystals of ErTe$_3$ shown in red and blue. An ideal Debye curve is drawn in black with $T_D = 160~$K.  The  CDW transitions  at $T_{CDW1}\approx 265~$K and $T_{CDW2}\approx 160$K produce no prominent critical signatures.  (b)  Resistivity measured on similar crystals rescaled so the room temperature resistivity is the same as in previous measurements. Blue for $a$ axis and red for $c$ axis.}
\label{cpres}
\end{figure}

\begin{figure}[ht]
\centering
\subfigure{\label{dif}\includegraphics[width=0.95\columnwidth]{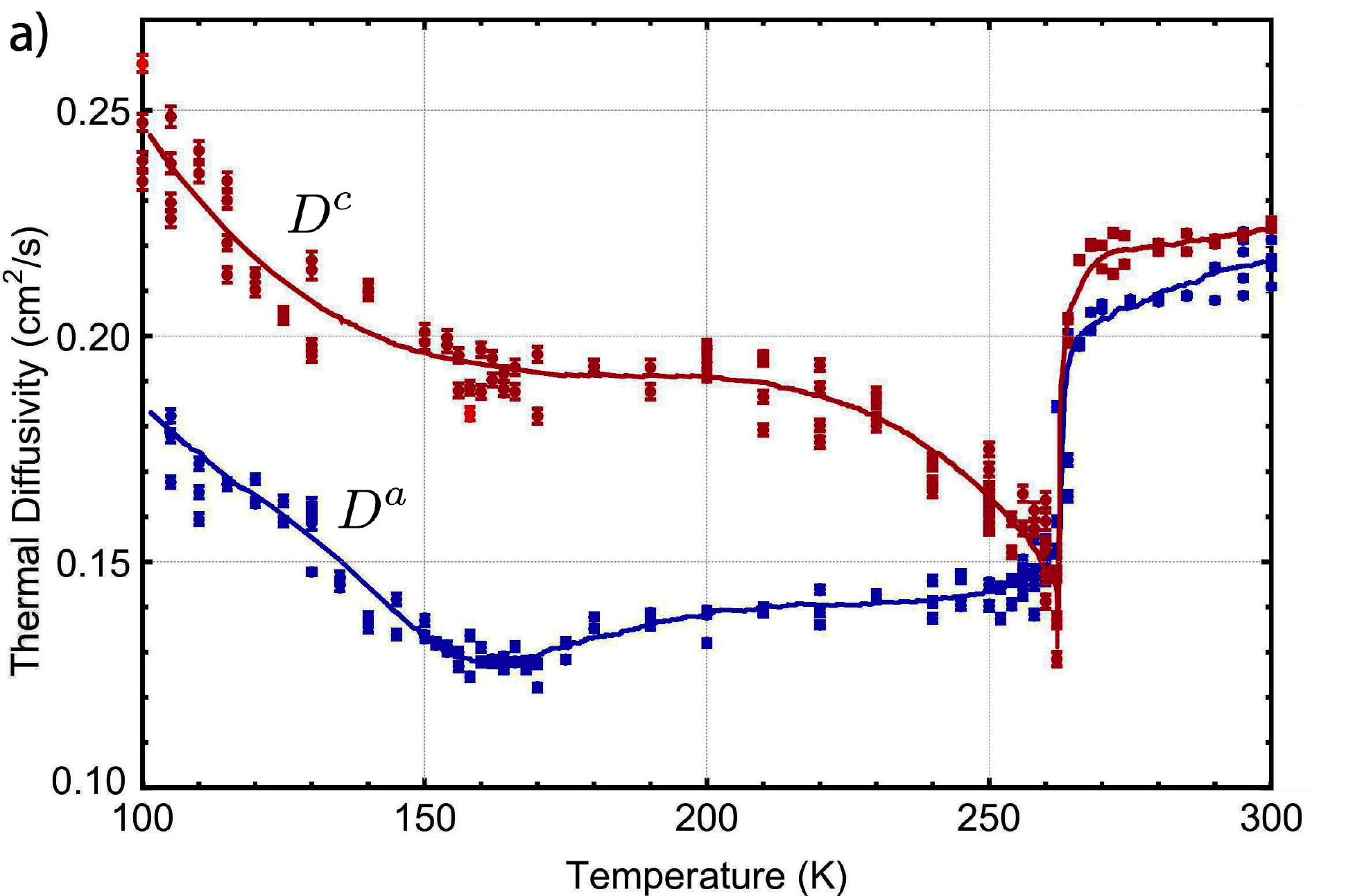} }
\includegraphics[width=0.95\columnwidth]{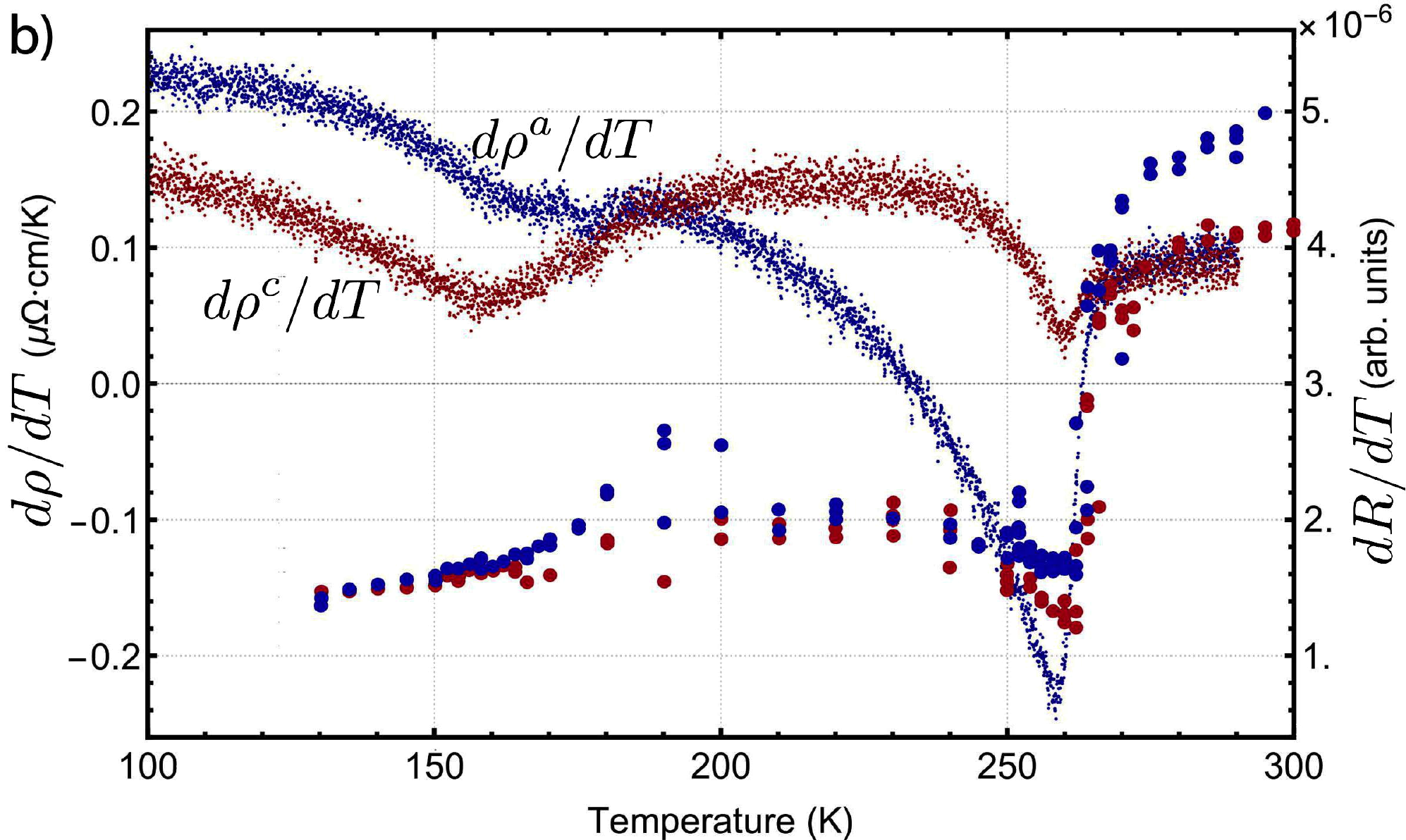} 
\caption{(a) Thermal diffusivity of ErTe$_3$ measured using a photothermal microscope. Scatter of data is primarily associated with different (although, nearby) positions on the crystal where the measurements were performed, and one-pixel variation in the relative distance between the heating and probing laser spots (see text). The two CDW transitions occur at $T_{CDW1}\approx 265~$K and $T_{CDW2}\approx 160~$K.  Blue for $a$ axis and red for $c$ axis.
(b) Temperature derivative of the reflectivity - $dR /dT$ (filled circles referenced to left axis) as a function of temperature shown in blue for $a$ axis and red for $c$ axis.  For comparison the dense points show the temperature derivative of the resistivity,  $d\rho /dT$, extracted from Fig.~\ref{cpres}(b) and referenced to the right axis. }
\label{diffref}
\end{figure}

Our primary result is that the thermal transport in a critical regime below $T_{CDW1}$ appears inconsistent with quasi-particle transport.  Specifically, by assuming separate electronic and phonon contributions to $\kappa$ and that $\kappa_{el}$ satisfies the WF law, one is forced to  infer an unphysically large depression of the lattice thermal transport in this regime. The regime in which we deduce the breakdown of WF law, and hence the breakdown of the quasiparticle concept, is found to be strongly asymmetric around $T_{CDW1}$, extending farther below $T_{CDW1}$  than above. In addition to this dramatic result, we also observe the following effects:  {\it i}) Similar to other strongly interacting CDW systems, large anomalies are observed in the temperature derivative of the resistivity and the reflectivity (Fig.~\ref{diffref}(b)), which, with the assumption that Fisher-Langer theory \cite{Fisher1968} applies, stands in sharp contrast to the small anomaly observed in heat capacity measurements \cite{SaintPaul2017}. {\it ii}) The behavior of the  various linear response tensors  near criticality (Figs.~\ref{cpres}(b) and~\ref{diffref}(a)) depends strongly on direction.   As a function of decreasing $T$, the resistivity along the $a$-direction, $\rho^a$, has a pronounced critical  singularity at $T_{CDW1}$ followed by  a broad maximum before continuing to drop  at lower temperatures, as has previously been discussed by Sinchenko {\it et al.}\cite{Sinchenko2014}.  On the other hand, the critical anomaly in $\rho^c$ at $T_{CDW1}$ is much weaker, and neither component shows any clear non-analyticity at $T_{CDW2}$.   
In contrast, the thermal diffusivity has a large and  sharp decrease at $T_{CDW1}$ along both the $a$- and $c$- directions,  followed by a relatively faster recovery along the $c$-direction.  In fact, as can be inferred from Fig.~\ref{diffref}(a), the thermal conductivity (see also Fig.~\ref{kappa} below) more closely resembles the temperature derivative of the resistivity, especially  the $a$-axis thermal diffusivity. {\it iii}) Thermal diffusivity in  both directions show marked increases below $T_{CDW2}$, which through suppression of this effect by weak Pd-intercalation are shown to be electronic in origin (Fig.~\ref{kappaPd} below).
 

\section{Experiment}

Samples of ErTe$_3$ were grown using a Te self-flux technique, which ensures purity of the melt, and produces large crystals with a high degree of structural order \cite{Ru2006}. The material is somewhat air sensitive, and crystals must be stored in an oxygen and moisture-free environment. Crystal orientation was determined with XRD.

The heat capacity of the single-crystal samples was measured using a relaxation time technique in a Quantum Design Physical Property Measurement System (PPMS). Crystals with a mass of approximately $5~$mg to $10~$mg and flat surfaces were selected for good thermal contact with the sample platform. Data were taken in zero applied field from $1.8$ to $300~$K. 

Measurements of resistance versus temperature were performed in a Janis Supertran-VP continuous flow cryostat. The resistivity in the $a$-$c$ plane was measured on thin rectangular crystals which had been cut with a scalpel and cleaved to expose a clean surface immediately before contacting. Crystals were cut in to bars such that current flows along the [101] axis, and contacts were attached to the surface in both longitudinal and transverse geometry, as described elsewhere \cite{Walmsley2017}. In this geometry, the longitudinal and transverse contacts simultaneously measure the sum of the resistivity components along the crystal axes $\rho^a+\rho^c$ and the in-plane resistivity anisotropy $\rho^a-\rho^c$ respectively. This technique minimizes uncertainty in the resistivity anisotropy arising from geometric effects relative to using two different samples to measure $\rho^a$ and $\rho^c$ separately. Measured resistivity values were then rescaled to match existing literature \cite{Ru2008,Walmsley2017} at room temperature while preserving the same anisotropy.

Thermal diffusivity and temperature derivative reflectivity ($dR/dT$) measurements were performed using a photo-thermal microscope as described in \cite{Zhang2017}. Further details associated with the present experiment are described in the Supplemental Material section\cite{SuppMat}. Using this apparatus, the thermal diffusivity is obtained directly, without the need to measure the thermal conductivity and specific heat separately. An advantage of this apparatus is the ability to measure the full in-plane anisotropy of the thermal diffusivity by orienting the pair of heating and probing laser spots at any arbitrary orientation with respect to the crystal axes. The mobility in the optics is further used for diagnostics of spatial uniformity of the thermal diffusivity and $dR/dT$ values.  Dimensions of the measured crystal were around $400~\mu$m by $300~\mu$m by $40~\mu$m. To minimize surface degradation from oxygen and water, samples were cleaved using Kapton tape immediately before being placed in the photo-thermal microscope and measured.

\section{Results}

The specific heat of two different ErTe$_3$ crystals over a wide temperature range is shown in Fig.~\ref{cpres}(a) along with an ideal Debye approximation fit with a Debye temperature $T_D=160~$K.  The data closely follow the Debye law both above and below each of the CDW transitions saturating at the Dulong-Petit value at high temperatures.  Previous measurements of the specific heat anomaly at the CDW transition ~\cite{SaintPaul2017} find that $\Delta c_p\approx 1 $J/mol$\cdot$K, below the resolution of the present measurements, and surprisingly of ``normal magnitude'' given the large value of $2\Delta_1/k_BT_{CDW1}\approx 15$ estimated via ARPES measurements\cite{Moore2010}. Thermal diffusivity of a different ErTe$_3$ crystal from the same growth batch is shown in Fig.~\ref{diffref}(a) along with a guide to the eye curve.  In contrast to the case of the specific heat, both CDW transitions produce  large anomalies in the diffusivity data. In particular, at $T_{CDW1}$ the diffusivity along both axes drops by over a third from $\sim 0.21~$cm$^2$/s to $\sim 0.14~$cm$^2$/s.

The sharp change in diffusivity can also be compared to the much broader change in electrical resistivity.  Resistivity data on the same-batch crystals is presented in Fig.~\ref{res}. The resistivity was rescaled to match existing literature as described previously. This data is similar in trend to previously measured data on RTe$_3$ crystals \cite{Ru2008}, in particular on ErTe$_3$ \cite{Walmsley2017,Straquadine2019}. The main feature in the resistivity data of this class of materials is the strong anomaly in the direction perpendicular to the primary CDW direction ($a$-axis), as  shown in Fig. \ref{res}. In contrast, the thermal diffusivity decreases along both axes. Moreover, below  $T_{CDW2}$, the thermal diffusivity increases along both axes while the resistivity is little affected.  However, the derivative of the resistivity $d\rho/dT$, shows effects of the CDW transitions that are more like those seen in the thermal diffusivity (Fig.~\ref{diffref}(a)). In all cases, there is a larger change at  $T_{CDW1}$ than at $T_{CDW2}$.   Finally, it is illuminating to compare the critical evolution of these various DC probes with the temperature derivative of the reflectivity at $h\nu \approx 1.5~$eV (820 nm wavelength) as shown in Fig. \ref{diffref}. The derivative of the reflectivity shows a large and sharp decrease at $T_{CDW1}$  while no anomaly is visible at $T_{CDW2}$.

The effect of purposefully introduced weak disorder on the temperature dependence of the thermal diffusivity is illustrated in  Fig.~\ref{wpd}, which shows a plot of the $a$ axis diffusivity of Pd$_{0.003}$ErTe$_3$ in green the data on pure  ErTe$_3$ already shown in Fig~\ref{dif}. $T_{CDW1}$ is slightly suppressed upon the intercalation, from $265$K to $250~$K, but  the sharp drop in the diffusivity at the transition looks just like the one in pure ErTe$_3$ (as shown in \cite{Fang2019}, no disorder effects appear in the $c$-direction for this weak disorder). However, below $T_{CDW2}$ (here around $130~$K) there is a striking intercalation induced difference; the pronounced upturn of the diffusivity in the pure material is eliminated with Pd intercalation.

\section{Discussion}


\subsection{Temperature derivative of the reflectivity and resistivity}
The resistivity of ErTe$_3$ has been measured before \cite{Ru2008,Walmsley2017} and the temperature dependence understood in terms of the band structure of the material \cite{Sinchenko2014}. Specifically, when the primary CDW forms along the $c$-axis, the resistivity begins to increase along the perpendicular $a$-direction. Likewise, at the formation of secondary CDW along the $a$-axis, there is a larger change in $d\rho/dT$ along the perpendicular $c$-direction.

The photothermal measurement used to extract thermal diffusivity information is a result of analyzing the phase delay in change of reflectivity from a probed point on the sample surface due to a propagating heat wave originating from a point-like source that is modulated at some frequency $\omega$ (see supplementary information \cite{SuppMat}). The amplitude of the reflected light $R(\nu)$, (where $\nu =c/\lambda$ is the frequency of the probing light with wavelength $\lambda$) detected at the probing point  can be shown to be proportional to $dR/dT$. The optical reflectivity of ErTe$_3$ was previously measured \cite{Pfuner2010,Hu2011} over the entire frequency range, exceeding the room-temperature plasma frequency, which depending on the global fit, is estimated in the range of $20,000~$cm$^{-1}$ ($500~$nm) \cite{Pfuner2010} to $47,000~$cm$^{-1}$ ($213~$nm) \cite{Hu2011}.  At our probing wavelength of 820 nm, and with a Drude scattering rate ($\Gamma_D\equiv1/\tau$) $\sim 20$ times smaller, a full Drude-Lorentz expression is needed to fit the experimental data in the whole frequency range  \cite{Pfuner2010,Hu2011}. Since the range of interband transition described by a set of Lorentz harmonic oscillators is temperature independent, it is reasonable to assume that much of the temperature-dependent component of the reduction in reflectivity comes from the temperature dependence of the relaxation time, which is strongly affected by scattering from CDW fluctuations  \cite{Pfuner2010}. In that case we can write $R(\nu)=R_0(\nu)+\Delta R(\nu\tau)$ (e.g., we may attempt to extend the Hagen-Rubens relation to near infrared corresponding to our probing light, $R(\nu)\simeq 1-2\sqrt{\nu\rho}$, where $\rho$ is the Drude resistivity), and thus $dR/dT\propto d\tau/dT$. 

In a seminal work, Fisher and Langer \cite{Fisher1968} showed that the leading (perturbative) effect of scattering of conduction electrons by classical (i.e. approximately static)  critical modes leads to the relation $dR/dT \propto c_{CDW}$, where $c_{CDW}$ is the specific heat associated with the  critical fluctuations in the neighborhood of  a finite $T$ phase transition. Examination of the temperature derivative of the resistivity data, particularly in the $a$-direction, indeed reveals what appears to be  a broadened discontinuity at $T_{CDW1}$, which is similar to the  behavior of the reflectivity.   This mean-field-like form is in agreement with the shape of the anomaly observed in direct measurements of specific heat in \cite{SaintPaul2017,SaintPaul2020}, although the relative strength of the anomaly is much weaker in those measurements (and essentially invisible in our Fig. 1a).  Despite their similar behaviors at  $T_{CDW1}$, at lower temperatures   $d\rho/dT$ and $dR/dT$ exhibit substantially different thermal evolutions.  The former, but not the latter recovers rapidly to values comparable to those the CDW transition\cite{Lazarevic2011}. Furthermore, near $T_{CDW2}$, $d\rho/dT$ shows a relatively weak but still clear critical anomaly, while the the effect of the second CDW transition is difficult to discern in $dR/dT$.   

\subsection{Thermal Conductivity}

More insight into the relation between electrical and thermal transport is obtained if we use the respective Einstein relations for these coefficients
\be
\sigma = \chi_{el} D_{el}; \ \ \ \ \ \ \kappa = c_pD_Q
\label{er}
\ee
where $\chi_{el}$ is the electronic compressibility, $c_p$ is its total specific heat, and $D_{el}$ and $D_Q$ are the electronic and heat diffusivities respectively. While $\chi_{el}$ is a response function of only the electron system, the specific heat of the material, particularly at high temperatures may be dominated by the lattice. In a simple kinetic approach where electrons and phonons transport heat in parallel channels we may write $\kappa = \kappa_{el}+\kappa_{ph} = c_{el}D_{el}+c_{ph}D_{ph}$, where $c_{el}$ and $c_{ph}$ are the electronic and lattice specific heats and $D_{el}$ and $D_{ph}$ are the respective diffusivities.
\begin{figure}[ht]
\centering
\includegraphics[width=1.0\columnwidth]{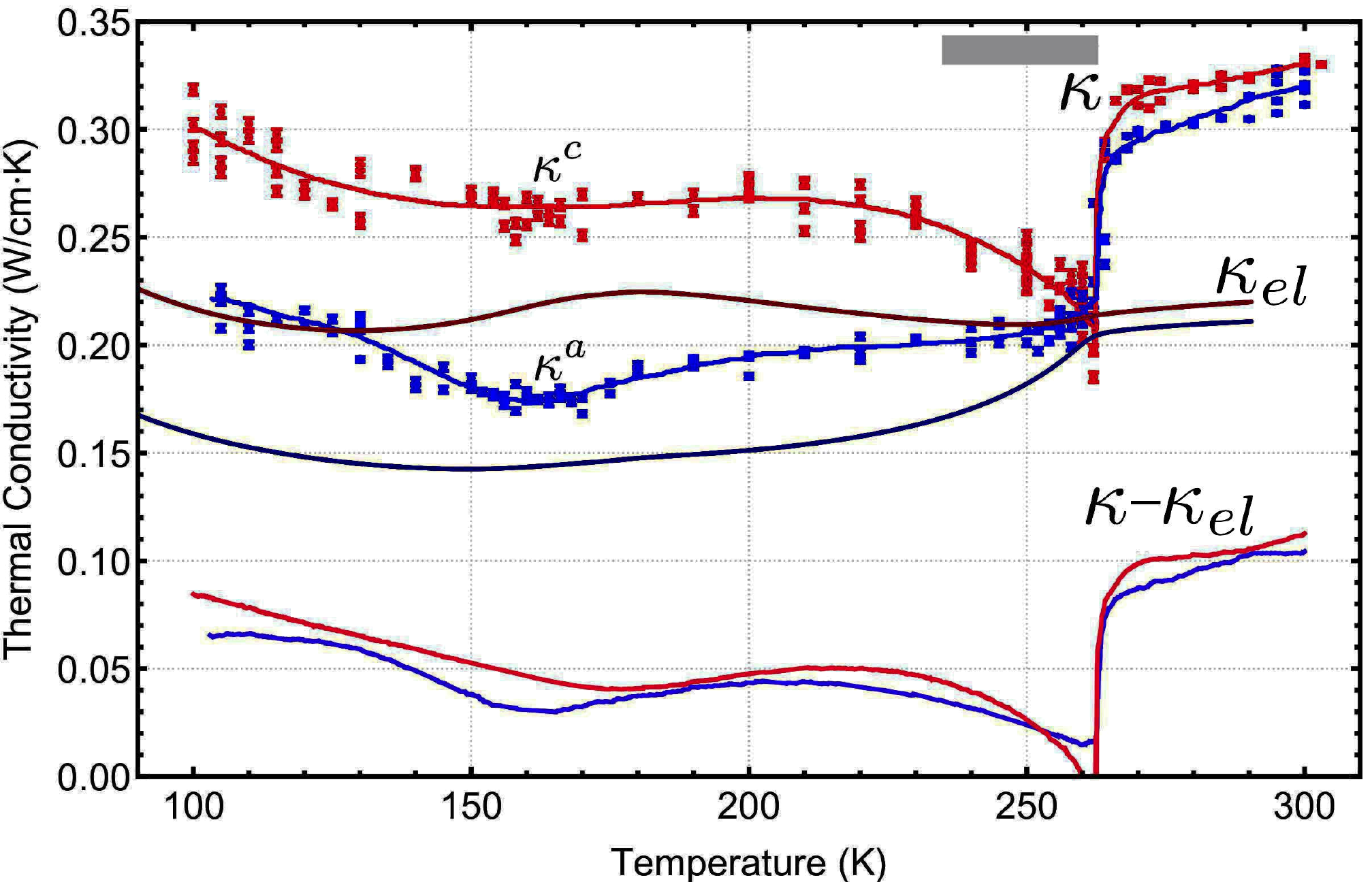} 
\caption{The total thermal conductivity, $\kappa$ (square symbols connected by a solid guide to the eye), the electronic component $\kappa_{el}$ computed from $\rho$ assuming the validity of the WF law (solid lines), and the ``non-electronic'' component, $\Delta\kappa$, (dashed lines) as a function of $T$.  The two lattice directions are indicated by red for c axis and blue for a.  The grey bar indicates a critical region where }
\label{kappa}
\end{figure}

Measuring the specific heat and thermal diffusivity, the total thermal conductivity can be calculated following Eqn.~\ref{er}. The resulting thermal conductivity is shown in Fig.~\ref{kappa} with blue and red showing the calculated thermal conductivity along the $a$ and $c$ axes respectively. A best-fit guide to the eye curve is drawn through the data points. Assuming that Wiedemann-Franz law holds, we can calculate the electronic thermal conductivity from the resistivity $\kappa_{el}=L_0T/\rho(T)$, which is also shown in Fig.~\ref{kappa}.  This allows us to define a ``non-electronic'' contribution $\Delta \kappa\equiv \kappa -\kappa_{el}$.  While it is conventional to identify $\Delta \kappa$ with an independent phonon contribution, $\Delta \kappa \leftrightarrow \kappa_{ph}$, it is apparent (as we will discuss below) that this is not plausible over much of the range of temperatures and especially in a region (indicated by the gray bar in Fig. 3)

We first consider the thermal conductivity around room temperature, above the CDW transitions. We note that the value of the total thermal conductivity is very high as compared to other chalcogenide-based CDW materials.  For example, $\kappa = 0.06~$W/cm$\cdot$K for TaSe$_3$ \cite{Zawilski2010}, $0.07~$W/cm$\cdot$K for NbSe$_3$ \cite{Yang2019}, $0.05~$W/cm$\cdot$K for (TaSe$_4$)$_2$I \cite{Kwok1989},  $0.1~$W/cm$\cdot$K for 2H-TaSe$_2$  \cite{Nunez1985},  $0.035~$W/cm$\cdot$K for HfTe$_5$ \cite{Zawilski2001}, or $0.08~$W/cm$\cdot$K at 370K for 1T-TaS$_2$ \cite{Nunez1985}, to give a few examples. By contrast, ErTe$_3$ exhibits $\sim 0.33$ W/cm-K at room temperature, which is more than 3 and up to 10 times larger thermal conductivity then those other compounds. However, using Wiedemann-Franz law and our measured resistivity to evaluate the electronic thermal conductivity, we obtain upon subtraction a value of $\Delta \kappa$ which is comparable in magnitude to the aforementioned materials. Moreover, in light of the considerably larger resistivities of these  other materials, the same Weidemann-Franz analysis is around 20\% to 25\% for most compounds, reaching 45\% for NbSe$_3$ nanowires \cite{Yang2019}.   Moreover, in all cases $\kappa$ is very weakly $T$ dependent in this range of $T$.  It is thus natural to identify $\Delta \kappa \approx \kappa_{ph}$ as an essentially independent phonon contribution to the thermal conductivity - as is commonly done.

While WF analysis is generally expected to work at temperatures comparable and above $T_D$, analysis of the CDW transition region, particularly the anomaly at $T_{CDW1}$, implies a catastrophic breakdown of this approach. Firstly, while on the basis of the WF law, one would expect the 
the critical anomaly in the total thermal conductivity to be weak, as it is in the resistivity, in fact it is pronounced and resembles the behavior of $d\rho/dT$.  (Note also that the specific heat anomaly at $T_{CDW1}$ is weak compared to the total heat capacity, primarily because of the high transition temperature where it is already in the Dulong-Petit regime). More dramatically, were we to use the WF law to subtract an electronic contribution to $\kappa$ in the critical regime, we would be forced to conclude that the lattice contribution $\Delta \kappa$ mysteriously vanishes, at least within $\sim30 $ degrees below $T_{CDW1}$ - the region indicated by the gray bar in Fig.~\ref{kappa}.  We are aware of no plausible physical mechanism that could produce such an effect.

Cooling down below $\sim240~$K, $\Delta \kappa$ reaches a value of order $0.05~$W/cm$\cdot$K, which is common to these type of materials as discussed above, and hence may again be loosely interpretted in terms of a parallel lattice contribution. Using simple kinetic theory, our measured specific heat and a typical longitudinal sound velocity of $\sim2.8\times 10^{5}$cm/sec \cite{SaintPaul2017,SaintPaul2020}, we obtain a mean free path of $\sim 35~$\AA~at $T=T_{CDW2}$, reduced from $\sim 80~$\AA~above $T_{CDW1}$. While below the primary CDW transition phonon mean free path might be expected to increase reflecting reduced phonon-electron scattering, CDW fluctuations in the transition region, and the CDW formation below that temperature could be additional sources of  phonon scattering.
 (However, by contrast, in the other chalcogenide-based CDW materials mentioned above, $\kappa$ is nearly constant with a slight tendency to increase with decreasing $T$ over the same range of $T$.) In the same temperature range the total and electronic thermal conductivities reach an anisotropic value of $\kappa^c/\kappa^a\approx\kappa_{el}^c/\kappa_e^a\approx 1.3 $, reflecting the effect of the primary CDW at $T_{CDW1}$. It is then interesting to note that this anisotropy is only weakly reduced below the secondary CDW transition at $T_{CDW2}$, where the primary effect is an increase in all components of the thermal transport. While the increase in the putative  lattice part below $T_{CDW2}$  could be due to further gapping of electronic states that decrease the phonon-electron scattering rate, the electronic increase in thermal conductivity simply reflects the increase of the mean free path of the   remaining itinerant electrons. This hypothesis can be checked by introducing additional electron disorder scattering with small concentration 
   of  intercolated Pd atoms ($\lesssim 1\%$).  At the same time, this level of intercalation does not seem to markedly change the carrier density \cite{Straquadine2019}.
\begin{figure}[ht]
\centering
\subfigure{\label{kepd}\includegraphics[width=0.95\columnwidth]{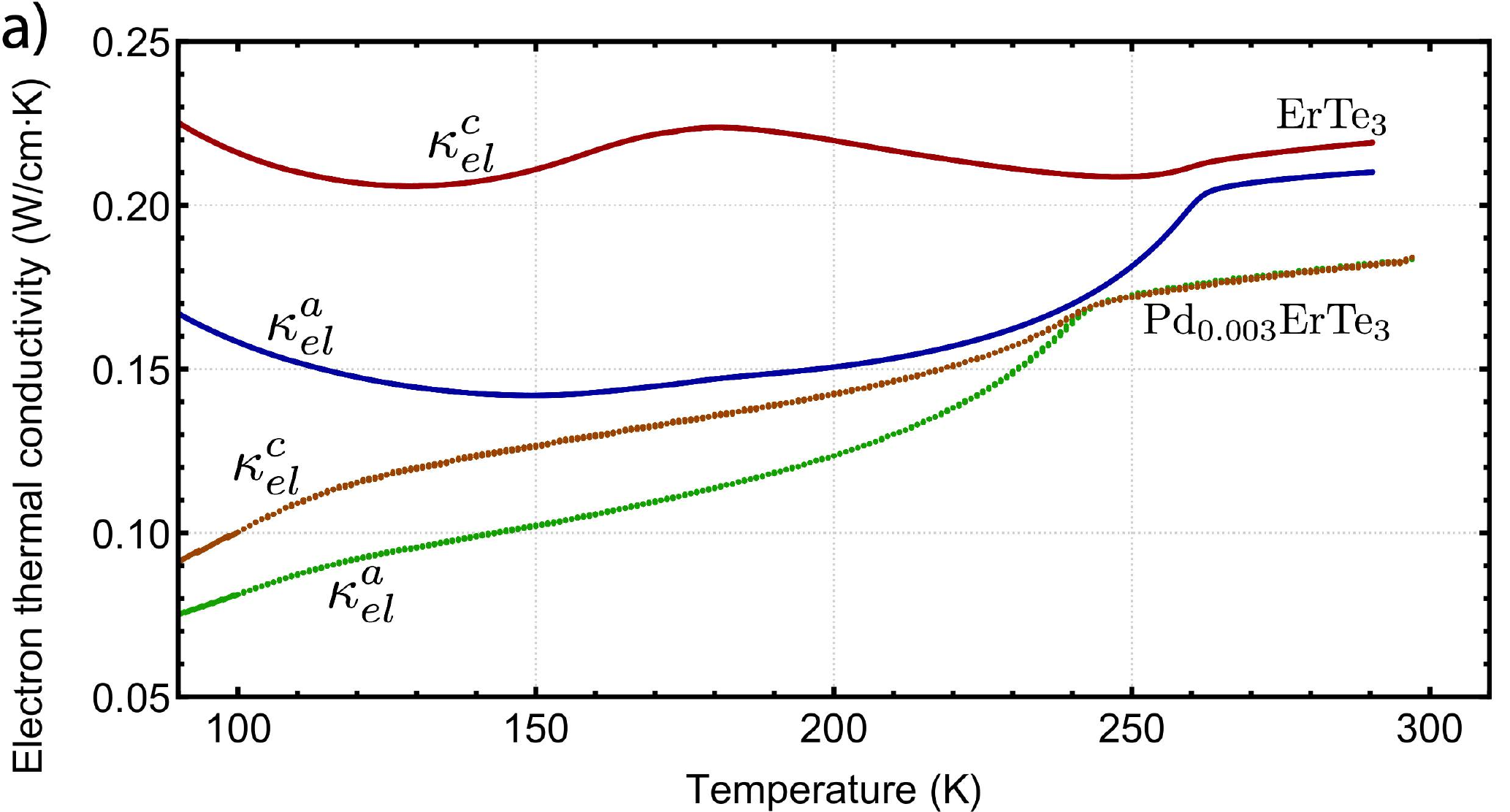} }
\subfigure{\label{wpd}\includegraphics[width=0.95\columnwidth]{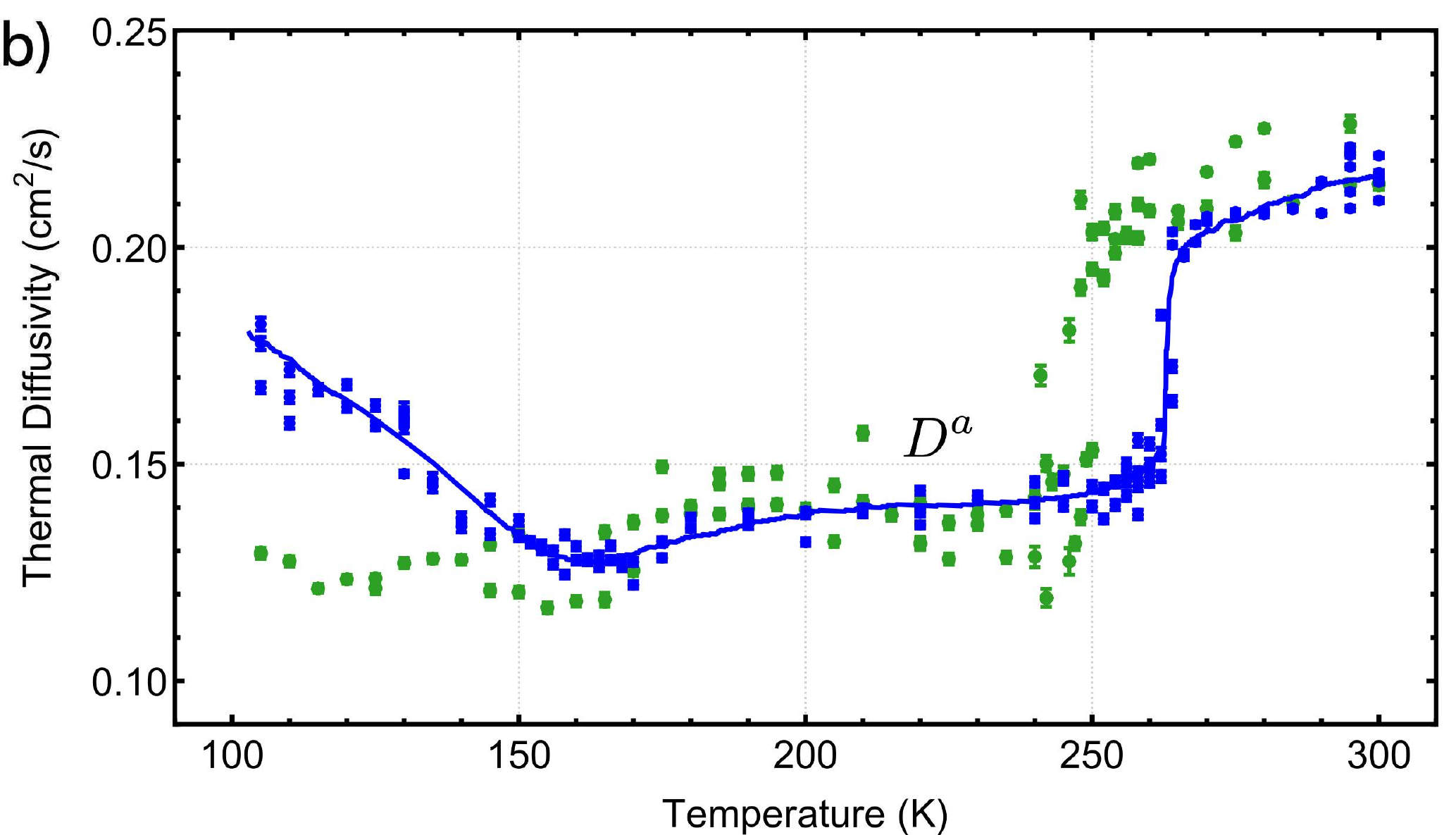} }
\caption{(a) Electronic portion of the thermal conductivity, $\kappa_{el}$  calculated from resistivity for ErTe$_3$ (top two curves) and Pd$_{0.003}$ErTe$_3$ bottom two curves.  (b)  Thermal diffusivity along the a axis of ErTe$_3$ (blue) and  of Pd$_{0.003}$ErTe$_3$ (green). Lines act as guide to the eye. }
\label{kappaPd}
\end{figure}

Figure~\ref{kappaPd} show the effect of $\sim 0.5\%$ Pd intercalation on resistivity and $\sim 0.3\%$ Pd intercalation effect on thermal conductivity. We first notice the respective decrease in the primary CDW transitions, which closely follows the phase diagram first introduced in \cite{Straquadine2019}. Focusing on the $a$-axis transport, where effects of disorder have been shown to the strong effect on the electronic component is observed, thermal diffusivity as shown in Fig.~\ref{wpd} did not change much below $T_{CDW1}$, but the characteristic increase increase below $T_{CDW2}$ is missing. The respective electronic component shown in Fig.~\ref{kepd} also show a decrease, both reflect the increase in disorder which scatter both, electrons and phonons. The incomplete gapping of electronic states may also affect the phonon-electron scattering \cite{Fang2019,Fang2020}.

\section{Broader Context}
In many cases, the transport properties of metals can be successfully understood on the basis of the response of weakly interacting elementary excitations - fermionic-quasi-particles and bosonic phonons.  In the last several decades, however, various transport regimes in certain ``highly correlated'' materials have been identified, in which the validity of such an approach has been called into question.  However,  it remains highly contravertial to what extent conventional quasi-particle ideas can be extended without fundamental changes in approach to strongly interacting regimes in which the quasiparticle identity is ``marginally'' maintained, or if entirely new paradigms (possibly involving  some form of ``non-Fermi liquid'' or novel fractionalized quasiparticles) are needed.  

One avenue of attack on this problem has been to investigate the way in which the usual quasiparticle picture can break down in the neighborhood of a quantum critical point.  However, even at classical (finite $T$) critical points, the existence of non-trivial critical exponents describing the behavior in the critical regime provides clear evidence that the critical modes themselves cannot have a quasi-particle description.  None-the-less, in many cases, as for instance in cases where the Fisher-Langer theory gives good account of transport anomalies, a treatment involving well-defined conduction electrons (and, presumably, phonons) weakly scattered by the critical modes, implies that the conventional mechanism of transport theory continues to apply even in the critical regime.  

The dramatic failure of such an approach to give an adequate description of the thermal transport in ErTe$_3$, most dramatically in a $\sim 30$K range of $T$ below $T_{CDW1}$, may be a first indication of a potentially simpler context to study the breakdown of the quasi-particle paradigm.  The discrepancies in the critical dependences of the thermal conductivity and the resistivity in this regime imply a complete breakdown of the WF law,  the existence of independent electronic quasiparticles and phonon modes, or both. In fact, the observed behavior may more adequately described as a strongly coupled electron-phonon critical `soup.'

CDW formation, arising from a variety of physical mechanisms, is a relatively common phenomenon in quasi-low-dimensional materials. Although thermal transport measurements have not been performed for all such materials, they do exist for many of the well-known canonical examples, and in none of these cases has such a dramatic violation of the WF law been deduced. This raises an associated question as to why the effect should be so pronounced in this particular material system given the relative ubiquity of CDW compounds (for a recent survey of CDW systems see e.g. \cite{SaintPaul2019}). A wider survey of related materials might yet reveal that this effect is in fact not unique to the rare earth tritellurides, but at least for now ErTe3 occupies a unique position among known CDW compounds and presents an entirely new opportunity to explore unconventional transport properties of strongly interacting metals.\\
\bigskip

\section{Acknowledgments}
We would like to thank Alan Fang for helpful discussions.  This work was supported by the Department of Energy, Office of Basic Energy Sciences, under contract no. DE-AC02-76SF00515. The photothermal apparatus was built using a grant from the Gordon and and Betty Moore Foundation through Emergent Phenomena in Quantum Systems Initiative Grant GBMF4529. JAWS was supported in part by an ABB Stanford Graduate Fellowship. AGS was supported in part by an NSF Graduate Research Fellowship (grant number DGE-1656518).

\bibliography{thermal}

\end{document}


	
\title{Anomalous thermal transport and violation of WF law through Charge Density Wave Transitions in ErTe$_3$}

\author{Erik Kountz}
\affiliation{Stanford Institute for Materials and Energy Sciences,\\
SLAC National Accelerator Laboratory, 2575 Sand Hill Road, Menlo Park, CA 94025}
\affiliation{Geballe Laboratory for Advanced Materials, Stanford University, Stanford, CA 94305}
\affiliation{Department of Applied Physics, Stanford University, Stanford, CA 94305}

\author{Jiecheng Zhang}
\affiliation{Stanford Institute for Materials and Energy Sciences,\\
SLAC National Accelerator Laboratory, 2575 Sand Hill Road, Menlo Park, CA 94025}
\affiliation{Geballe Laboratory for Advanced Materials, Stanford University, Stanford, CA 94305}
\affiliation{Department of Applied Physics, Stanford University, Stanford, CA 94305}

\author{Joshua A. W. Straquadine}
\affiliation{Stanford Institute for Materials and Energy Sciences,\\
SLAC National Accelerator Laboratory, 2575 Sand Hill Road, Menlo Park, CA 94025}
\affiliation{Geballe Laboratory for Advanced Materials, Stanford University, Stanford, CA 94305}
\affiliation{Department of Applied Physics, Stanford University, Stanford, CA 94305}

\author{Anisha G. Singh}
\affiliation{Stanford Institute for Materials and Energy Sciences,\\
SLAC National Accelerator Laboratory, 2575 Sand Hill Road, Menlo Park, CA 94025}
\affiliation{Geballe Laboratory for Advanced Materials, Stanford University, Stanford, CA 94305}
\affiliation{Department of Applied Physics, Stanford University, Stanford, CA 94305}

\author{Maja D. Bachmann}
\affiliation{Stanford Institute for Materials and Energy Sciences,\\
SLAC National Accelerator Laboratory, 2575 Sand Hill Road, Menlo Park, CA 94025}
\affiliation{Geballe Laboratory for Advanced Materials, Stanford University, Stanford, CA 94305}
\affiliation{Department of Applied Physics, Stanford University, Stanford, CA 94305}

\author{Ian R. Fisher}
\affiliation{Stanford Institute for Materials and Energy Sciences,\\
SLAC National Accelerator Laboratory, 2575 Sand Hill Road, Menlo Park, CA 94025}
\affiliation{Geballe Laboratory for Advanced Materials, Stanford University, Stanford, CA 94305}
\affiliation{Department of Applied Physics, Stanford University, Stanford, CA 94305}

\author{Steven A. Kivelson}
\affiliation{Stanford Institute for Materials and Energy Sciences,\\
SLAC National Accelerator Laboratory, 2575 Sand Hill Road, Menlo Park, CA 94025}
\affiliation{Geballe Laboratory for Advanced Materials, Stanford University, Stanford, CA 94305}
\affiliation{Department of Physics, Stanford University, Stanford, CA 94305}

\author{Aharon Kapitulnik}
\affiliation{Stanford Institute for Materials and Energy Sciences,\\
SLAC National Accelerator Laboratory, 2575 Sand Hill Road, Menlo Park, CA 94025}
\affiliation{Geballe Laboratory for Advanced Materials, Stanford University, Stanford, CA 94305}
\affiliation{Department of Applied Physics, Stanford University, Stanford, CA 94305}
\affiliation{Department of Physics, Stanford University, Stanford, CA 94305}

\setcounter{figure}{0}
\renewcommand{\thefigure}{S\arabic{figure}}%

\newpage
\newpage


\begin{center}
{\Large Supplemental Material for}\\
{\bf Anomalous thermal transport and violation of WF law through Charge Density Wave Transitions in ErTe$_3$}\\
{\normalsize  Erik D. Kountz,$^{1,2,3}$ Jiecheng Zhang,$^{1,2,3}$ Joshua A. W. Straquadine,$^{1,2,3}$ Anisha G. Singh,$^{1,2,3}$ Maja D. Bachmann,$^{1,2,3}$ Ian R. Fisher,$^{1,2,3}$ Steven A. Kivelson,$^{1,2,4}$ and
Aharon Kapitulnik$^{1,2,3,4}$}\\
\end{center}

\bigskip
\section*{ MATERIALS, METHODS AND ADDITIONAL INFORMATION}


\subsection*{Single Crystals Growth}

ErTe$_3$ and Pd$_{0.003}$ErTe$_3$ single crystals were prepared by self-flux with a typical size of $0.4\times0.3\times0.04$ mm$^3$ as described in Ref.~\cite{Ru2006}. All crystals measured of a single intercalation value were same-batch crystals.

\subsection*{Resistivity measurements}

The resistivity in the $a-c$ plane was measured on thin rectangular crystals which had been cut with a scalpel and cleaved to expose a clean surface immediately before contacting. Crystals were cut such that current flows along the [101] axis, and contacts were attached to the surface in the transverse geometry\cite{Walmsley2017}. In this geometry, the sum of the resistivity components along the crystal axes $\rho_a+\rho_c$ and the in-plane resistivity anisotropy $\rho_a-\rho_c$ are measured simultaneously within the same crystal. The largest source of error in such resistivity measurements is the magnitude of the resistivity resulting from the uncertanty in sample dimensions. Previous measurements of resistivity in the $a-c$ plane \cite{Ru2008,Walmsley2017} were used to as a room temperature reference to rescale the resistivity presented in this letter. The raw resistivity data measured was lower than previous measurements. If the data was not rescaled to correct for uncertanty in sample size, the resulting electrical thermal conductivity would be larger and the phononic thermal conductivity would be smaller than presented. Differential resistivity was calculated by taking the numerical derivative of the resistivity data. Data was smoothed using a mean and median filter.

\subsection*{Thermal Diffusivity Measurements}

\subsubsection*{Principles of the Photothermal Apparatus}

\begin{figure}[ht]
	\centering
	\includegraphics[width=1.0\columnwidth]{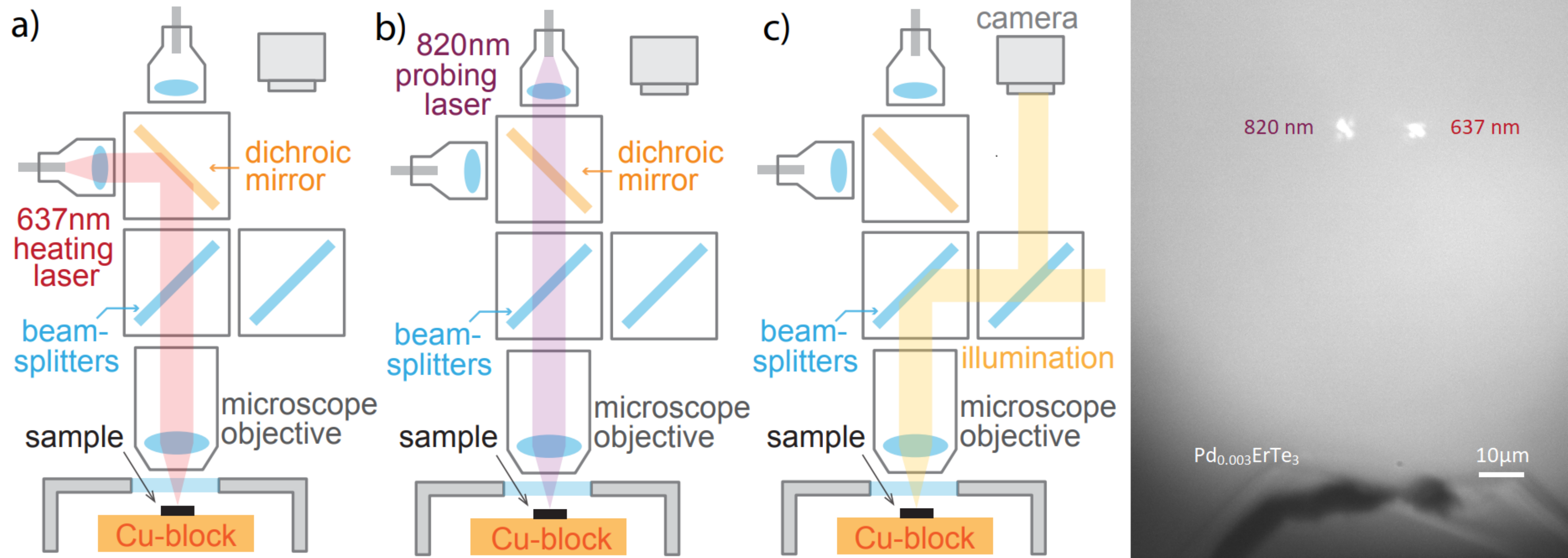}
	\caption{\label{setup}(color online) The schematic shows the optical paths of the setup. (a) Path of the heating laser (b) Path of the probing laser. The reflected light traverses the same path before gathered by a photodetector. (c) Path of illumination light source and camera vision. (d) Focused laser spots at typical measurement separation distance. The screenshot is taken on a just cleaved Pd$_{0.003}$ErTe$_3$ sample surface for visual interest, where atomically flat terraces in the bottom of the image can be seen. The surface of crystals measured in the main text are mostly featureless.}
\end{figure}

For the high resolution thermal diffusivity measurements we use a home-built photothermal microscope.  The microscope views the sample through a sapphire optical window in a cryostat, with the sample mounted to a cold finger just under the window.  A schematic is shown in Fig.~\ref{setup}. A heating laser at 637 nm and a probing laser at 820 nm are focused onto the sample surface by the microscope objective.  The focused spots have Gaussian radii of approximately $1~\mu$m and $2~\mu$m, respectively, due to the diffraction limit of different wavelengths, and can be moved independently over the sample surface.  A camera allows us to observe the sample surface nearby, align the spots in a particular orientation with respect to the crystal, and determine the distance between the spots.  

The output power of the heating laser is modulated with a sinusoidal profile $P(t)=P_0 [1 + \sin(\omega_0 t)]$. The modulation frequency $\omega_0/2\pi$ has a typical range of $1.5~$kHz - $25.5~$kHz, much slower than the microscopic equilibration on the order of picoseconds. This means that the parameters extracted are all within the in the DC limit of linear response, and that the dependency on the modulation frequency can be neglected. The probing laser is aimed at a spot a small distance (typically $10-20~\mu$m) away from the heating laser. The reflected light from the probing laser is diverted by an optical circulator and fed into a photodetector. The AC component of the photodetector signal is then fed to a lock-in amplifier referenced to the laser modulation and the amplitude and phase are measured. The DC component is used as a gauge to make sure the lasers are focused.


\subsubsection*{Measuring Thermal Diffusivity}
The diffusive transport of heat is governed by the diffusion equation
\begin{equation}
\label{diffusioneq}
	\frac{\partial\,\delta T(t,\vec{r})}{\partial t}
-D\nabla^2\,\delta T(t,\vec{r})
=\frac{q(t,\vec{r})}{c}
\end{equation}
where $\delta T$ is the temperature disturbance above the ambient temperature $T$, $\vec{r}=(x_1, x_2, x_3)$ is the spherical radial coordinate given in terms of the euclidean principal axes $x_i$, $q$ is the absorbed power density, $c$ is the volumetric specific heat capacity, $D \equiv \kappa / c$ is the thermal diffusivity, and $\kappa$ is the thermal conductivity.  Note that $c$ and $D$ are themselves functions of $T$, but in the limit of weak heating $\delta T \ll T$, we make the approximation $c(T+\delta T)\approx c(T)$ and $D(T+\delta T)\approx D(T)$. The temperature disturbance from both lasers is estimated to be $\lesssim 1~$K through out the temperature range, so the previous approximation is valid.

The modulated power of the heating laser causes ripples in the temperature profile at the sample surface, the resulting change in reflectivity can then be picked up by the probing laser.  It is useful to write the response in frequency space $\tilde{\delta T}(\omega,\vec{r})$, where
\begin{equation}
\delta T(t,\vec{r}\,)=\int_{-\infty}^\infty\tilde{\delta T}(\omega,\vec{r})\exp(-i\omega t)d\omega 
\end{equation}
We model the focused heat source as a point source, $q(t,\vec{r}\,)= P_0 e^{-i\omega t}\delta^3(\vec{r})$.  This approximation is valid as long as the distance from the heating spot is much larger than the spot radius.  In a semi-infinite isotropic system, the temperature profile is spherically symmetric and takes the form
\begin{equation}
	\tilde{\delta T}(\omega,r)=\underbrace{\frac{P_0}{\kappa}
	\frac{1}{r}\exp \bigg(-\sqrt{\frac{\omega}{2D}}r\bigg)}_{\text{amplitude}}
		\underbrace{\exp\bigg(-i\sqrt{\frac{\omega}{2D}}r\bigg)}_{\text{phase}}
			\label{diffsol}
\end{equation}

\begin{figure}[ht]
	\centering
	\includegraphics[width=01.0\columnwidth]{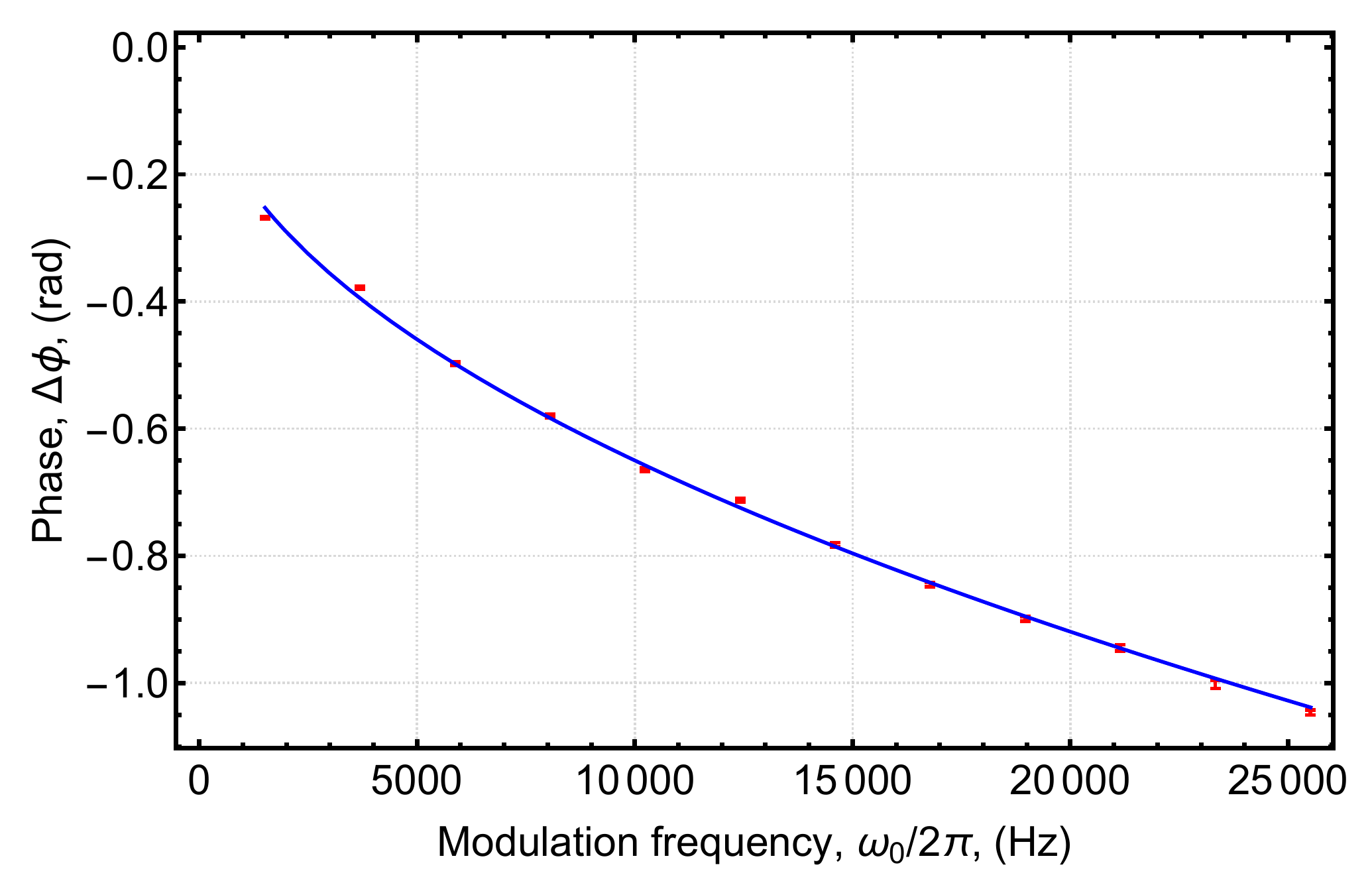}
	\caption{\label{phase}A typical phase delay as a function of heating laser modulation frequency, $\phi(\omega_0)$(red dot and error bars) obtained by sweeping $\omega_0$. This particular set is measured on ErTe$_3$ along the $a$ axis at a separation distance of $16.8~\mu$m at $300~$K. The blue line is a fit using the form in Eqn.~\ref{diffsol}.}
\end{figure}

We vary the separation distance to verify the semi-infinite 3D system assumption and the small spot assumption, and we vary the heating power to verify the weak heating assumption. Our measurement gives us the response at the modulation frequency $\omega_0$. Because the separation distance between the lasers spots was measured using camera vision and each individually mounted sample comes at a small random tilt, a systematic error on the order of 6\% is associated to each set of measurements.  Although both the amplitude and the phase of the solution carry information about $D$, in actual measurements factors such as mechanical vibrations, fluctuations in the laser power, surface imperfections, and the temperature dependence of the differential reflectivity $dR/dT$ affect the amplitude of the reflectivity oscillation.  These factors only enter the solution by multiplying real numbers, and thus do not affect the phase of the signal, which is therefore the more robust probe. We obtain $D$ by fitting the phase delay $\phi$ between the source and the response signals as a function of $\omega$ at fixed $r$: $D = \omega r^2 / 2 \phi^2$. A typical fit is shown in Fig.~\ref{phase}. All measurements of diffusivity were taken by measuring the phase delay at 12 evenly spaced frequencies between $1.5~$kHz and $25.5~$kHz and fitting for diffusivity using Eqn.~\ref{diffsol}.
We check the homogeneity of the crystals by repeating measurements at different positions on the surface, and check the isotropy/anisotropy by rotating the relative orientation between the laser spots.


\bibliography{thermal}